\documentclass[aps,pra,groupedaddress,showpacs,twocolumn]{revtex4-1}
\usepackage{graphicx}
\usepackage{amsmath}
\usepackage{bbm}
\usepackage{xspace}
\usepackage{color}
 \definecolor{eacol}{cmyk}{0,0.81,0,0}
\newcommand{\bareps}{\bar\epsilon}

\newcommand{\ie}{i.\thinspace{}e.\@\xspace}

\newcommand{\ve}[1]{{\bf #1}}
\newcommand{\mat}[1]{\mathsf{#1}}

\newcommand{\eq}[1]{Eq.\thinspace{}(\ref{#1})}
\newcommand{\eqq}[2]{Eqs.\thinspace{}(\ref{#1}) and (\ref{#2})}

\newcommand{\fig}[1]{Fig.\thinspace{}\ref{#1}}

\newcommand{\fc}[1]{({#1})}
\newcommand{\figc}[2]{Fig.\thinspace{}\ref{#1}\thinspace{}\fc{#2}}

\newcommand{\Fig}[1]{Figure~\ref{#1}}

\newcommand{\Tr}{\mbox{Tr}}

\DeclareMathOperator{\diag}{diag}
\newcommand{\nag}{{\phantom{\dag}}}

\begin{document}

%\preprint{}

\title{Excitations in disordered bosonic optical lattices}


\author{Michael Knap}
\email[]{michael.knap@tugraz.at}
\affiliation{Institute of Theoretical and Computational Physics, Graz University of Technology, A-8010 Graz, Austria}
\author{Enrico Arrigoni}
\affiliation{Institute of Theoretical and Computational Physics, Graz University of Technology, A-8010 Graz, Austria}
\author{Wolfgang von der Linden}
\affiliation{Institute of Theoretical and Computational Physics, Graz University of Technology, A-8010 Graz, Austria}
%\thanks{}


\date{\today}

\begin{abstract}
Spectral excitations of ultracold gases of bosonic atoms trapped in one-dimensional optical lattices with disorder are investigated by means of the variational cluster approach applied to the Bose-Hubbard model. Qualitatively different disorder distributions typically employed in experiments are considered. The computed spectra exhibit a strong dependence on the shape of the disorder distribution and the disorder strength. We compare alternative results for the Mott gap obtained from its formal definition and from the minimum peak distance, which is the quantity available from experiments.
\end{abstract}

%64.70.Tg: Quantum phase transitions (for quantum Hall effects aspects, see 73.43.Nq in electronic structure of surfaces, interfaces, thin films, and low dimensional structures)
%67.85.De: Dynamic properties of condensates; excitations, and superfluid flow
%03.75.Kk: Dynamic properties of condensates; collective and hydrodynamic excitations, superfluid flow
%71.36.+c: Polaritons (including photon-phonon and photon-magnon interactions)
%73.43.Nq       Quantum phase transitions (see also 64.70.Tg Quantum phase transitions in equations of state, phase equilibria and phase transitions)
%64.70.-p       Specific phase transitions
%42.50.Ct       Quantum description of interaction of light and matter; related experiments
%05.30.Jp       Boson systems (for static and dynamic properties of Bose-Einstein condensates, see 03.75.Hh and 03.75.Kk; see also 67.10.Ba Boson degeneracy in quantum fluids)
%63.50.-x       Vibrational states in disordered systems
%\keywords{}

\maketitle

%\section{\label{sec:introduction}Introduction}

%\textit{Introduction.---}
Interacting many-body systems with disorder are fascinating and
challenging from both the experimental as well as the theoretical
point of view. Understanding disordered bosonic systems has been of
great interest ever since the pioneering works on the Bose-Hubbard
(BH) model \cite{fisher_boson_1989}, which describes strongly
interacting lattice bosons. Originally, the disordered BH model has
been used to approximately describe various condensed matter systems,
such as superfluid helium absorbed in porous media
\cite{Crowell_1995,Csathy_2003}, superfluid films on substrates
\cite{Haviland_1989}, and Josephson junction arrays
\cite{vanDerZant_1996}. However, seminal experiments on ultracold
gases of atoms trapped in optical lattices shed new light on
interacting bosonic many-body systems, as these systems provide a
\textit{direct} experimental realization of the BH model
\cite{jaksch_cold_1998,bloch_many-body_2008}. Intriguingly, these
experiments {allow}
to observe quantum many-body phenomena, such as
the quantum phase transition {from a superfluid  to a Mott state}
\cite{greiner_quantum_2002}. 
The condensate of atoms can be driven across this phase transition by
gradually increasing the laser beam intensity, which is directly
{related} to the depth of the potential wells. There is a large
experimental control over the system parameters such as the particle
number or lattice depth, and in addition the parameters are tuneable
over a wide range. While optical lattices provide a very clean
experimental realization of strongly correlated lattice bosons, they
can be used to study disordered systems on a very high level of
control as well. Disorder can be added to the regular optical lattice
by several techniques, such as by superposing additional optical
lattices with shifted wavelength and beam angles
\cite{aspect_anderson_2009, Fallani_2007,Roth_2003, Damski_2003,Hild_2006,Hild_2009}, laser speckle fields
\cite{pasienski_disordered_2010,White_2009,Lye_2005, shrestha_correlated_2010}, or including atoms of a different species
acting as impurities \cite{Ospelkaus_2006}.

%potential
\begin{figure}
        \centering
        \includegraphics[width=0.42\textwidth]{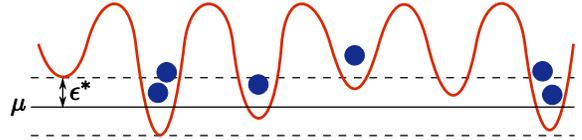}
        \caption{(Color online) The periodic optical lattice potential is locally modified by disorder. In the illustration the disorder is bounded by $\epsilon^*$ corresponding to a situation obtained by the superposition of two incommensurate optical lattices. }%
        \label{fig:potential}
\end{figure}

The disordered BH model has been widely investigated in the
literature. Most of the work has been devoted to study the phase
transitions occurring in the disordered BH model \cite{Scalettar_1991,Krauth_1991,Sheshadri_1995,Freericks_1996,Svistunov_1996,Kisker_1997,Herbut_1997,Prokofev_2004,Balabanyan_2005,Buonsante_2007,Weichman_2008,Pollet_2009,Gurarie_2009,Bissbort_2009,krueger_anomalous_2009,krueger_two_2010}. At zero temperature
the ground-state phase diagram depending on the chemical potential, the tunneling probability of the particles and the disorder strength, 
consists of three different phases: the
Mott {insulating,} the superfluid, and the Bose glass
phase. The first two phases are already present in the pure BH model,
{while} the latter is a distinct feature of the disordered system. 
 The Bose glass phase is characterized by being gapless and
  compressible, however, due to disorder the phase coherence does not extend
  over the entire system in contrast to the superfluid phase.
{While the phase diagram of the disordered BH model has been extensively studied,}
up to now
spectral properties  have not yet been
investigated theoretically, even though they are experimentally
accessible by Bragg spectroscopy, which allows to extract the wave
vector dependent excitation energies of the system
\cite{clement_2009,fabbri_2009,ernst_probing_2010}. For low-energy
excitations (typically $\omega/2\pi \lesssim 10\text{kHz}$) the
response of the Bragg spectroscopy is ascribed to the structure
factor, which corresponds to density fluctuations, whereas, for high 
excitation energies ($\omega/2\pi \gtrsim 30\text{kHz}$) the
single-particle spectral function {can be extracted}
\cite{clment_multi-band_2009,clment_bragg_2010}.
{In this case,}
atoms excited {into} high-energy bands are not expected to interact with the
lower bands. Therefore, the
measured spectrum describes a convolution (without vertex corrections)
  of the density of states of the lower occupied with the upper
  unoccupied
 band. From this convolution the spectral function
of the lower strong-correlated bands can be extracted \cite{clment_multi-band_2009}.

In the present paper we study in detail the
spectral properties of
bosonic atoms in a disordered optical lattice, modeled by the one-dimensional
BH model 
in which, for simplicity, the harmonic trap potential has been neglected.
We focus on the strongly correlated regime and evaluate spectral properties for the hopping strength to the on-site interaction ratio of size $t/U=0.05$. This ratio corresponds to an optical lattice depth of $11\,E_R$ , where $E_R$ is the recoil energy \cite{zwerger_mott_2003}. To evaluate this depth we considered laser beams with wave length $\lambda=830\,nm$ and the scattering length of rubidium $a_s \approx 5 \, nm$. 
In particular, we focus on the similarities and differences between
the two common  experimental methods used
 to induce disorder: (i)
superposition of two laser fields with incommensurate wavevectors on
the one hand, and (ii)
the addition of a laser speckle field on the other hand.
The disordered BH model and the disorder distributions generated by the two experimental methods are introduced in Sec.~\ref{sec:model}. 

\begin{figure*}
        \centering
        \includegraphics[width=0.9\textwidth]{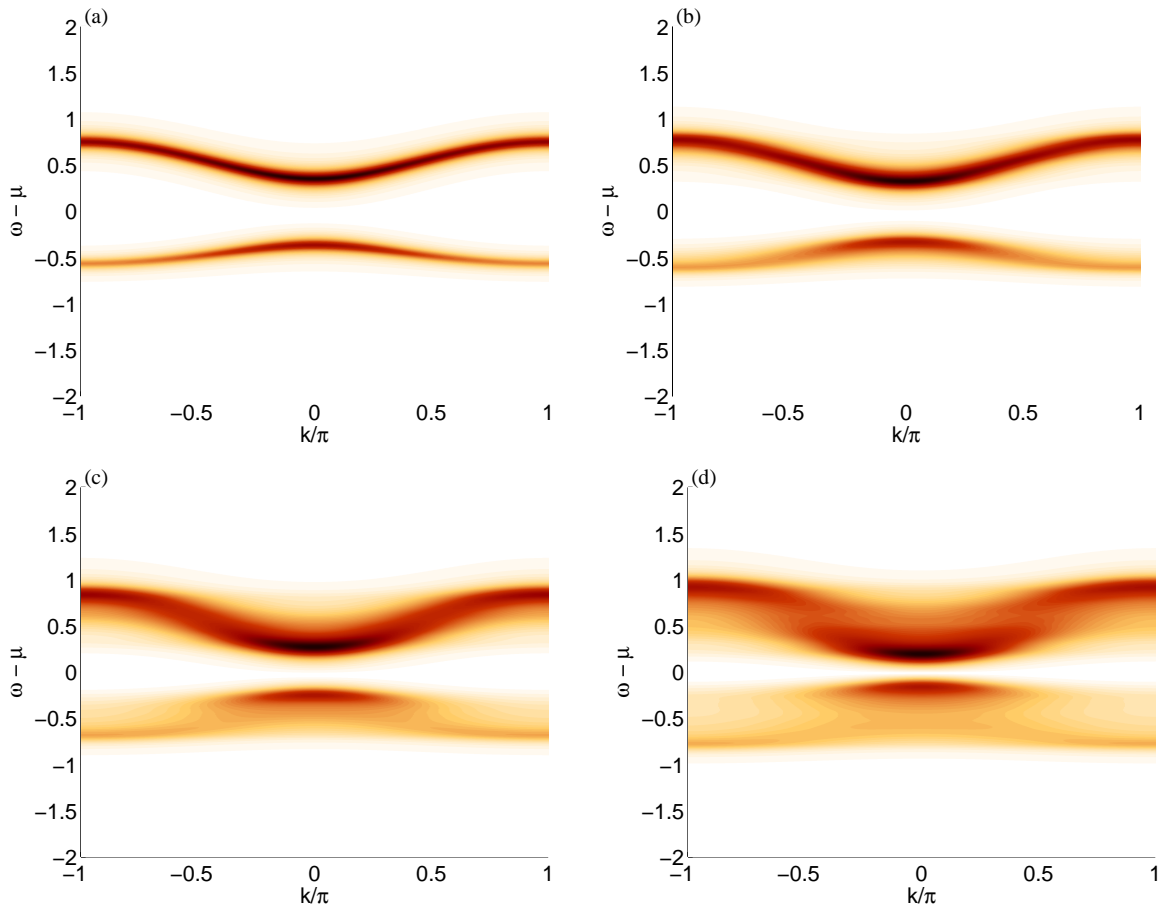}
        \caption{(Color online) Spectral function $A_p(\ve
          k,\,\omega)$ for disorder generated by {superposition
            of} incommensurate
          optical lattices. The parameters are $t=0.05$, $\mu=0.45$,
          and \fc{a} $\epsilon^*=0.0$, \fc{b} $\epsilon^*=0.1$, \fc{c}
          $\epsilon^*=0.2$, and \fc{d} $\epsilon^*=0.3$. }%
        \label{fig:sf}
\end{figure*}

We investigate the spectral properties of the disordered BH model numerically using the variational cluster approach \cite{potthoff_variational_2003}. In this work we present the first application of the variational cluster approach on disordered, interacting many-body systems. A description of the method is given in Sec.~\ref{sec:method}. Our results on the excitations of the disordered BH model are presented in Sec.~\ref{sec:results}. 
The crucial point about our investigations is that the two experimental approaches described previously
produce disorder with rather
different distributions leading, in turn, to different distributions of the spectral weight, as we
show  in Figs.~(\ref{fig:sf}) and (\ref{fig:sfExp}). In addition, we present two results for the Mott gap.
The first one, termed $\Delta$,
 is obtained from the smallest excitation energy.
 This is, in principle, the formally correct version, but it
 is hard to determine experimentally.
The second one, termed $\Delta^{mp}$, is {evaluated}
from the minimum peak distance in the spectral weight, which is the
accessible quantity in experiments.
The two quantities
  exhibit a different behavior as shown in Fig.~\ref{fig:gap}.
Finally, in Sec.~\ref{sec:conclusion}, we conclude and summarize our work.

\section{\label{sec:model}Disordered Bose-Hubbard model}
%\textit{Disordered Bose-Hubbard model.---}
The Hamiltonian of the BH model with {on-site} disorder \cite{fisher_boson_1989} is given by
\begin{equation}
 \hat{H}=-t \sum_{\left\langle i,\,j \right\rangle} b_i^\dagger \, b_j^\nag
+ \frac{U}{2} \sum_i \hat{n}_i\left(\hat{n}_i-1 \right) + \sum_i (\epsilon_i - \mu)\hat{n}_i \; \mbox{,}
 \label{eq:bhm}
\end{equation}
where the operators $b_i^\dagger$ ($b_i^\nag$) create (annihilate)
bosonic particles at lattice site $i$.
The first sum in \eq{eq:bhm} is restricted to nearest-neighbor
sites. The hopping strength is denoted as $t$, $U$ is the on-site
repulsion and $\mu$ stands for the chemical potential, which controls the
total particle number $\hat{N}_p=\sum_i \hat{n}_i = \sum_i b_i^\dagger
\, b_i^\nag$. The site-dependent random variable $\epsilon_i$ introduces disorder. 
 In the following calculations and discussions we use the
local interaction $U$ as unit of energy.

 One {sample-}configuration of disorder is denoted as
$\eta = (\epsilon_1,\,\epsilon_2 \, \ldots \, \epsilon_N)$, where $N$
is the number of lattice sites.  
The disorder modifies locally the depth of the optical lattice, see \fig{fig:potential} for illustration.
{Random} disorder is distributed
according to {a given} probability distribution function {(pdf)}
$p(\eta)$, which has to satisfy the normalization condition $\int
p(\eta)\,d\eta = 1$. The average of a quantity $X_\eta$ with respect
to the pdf $p(\eta)$ is given by $X_p \equiv \langle X \rangle_p =
\int p(\eta)\,X_\eta \, d\eta$. We consider $\epsilon_i$ as
identically and independently distributed random variables leading to $ p(\eta) = \prod_{i=1}^N
q(\epsilon_i)$, where $q(\epsilon_i)$ is a pdf describing the disorder
generated in the experiment. Experimentally, there are two main ways of introducing disorder,
corresponding  to two different pdf's.
On the one hand,
superposing two incommensurate optical lattices
{yields}
a shifted $\beta$ distribution as demonstrated in Appendix \ref{app:a}.
The distribution is bounded
by a maximum disorder strength $\epsilon^*$.
This approach is used, for example, in the setup by
 Fallani \textit{et al.}
\cite{Fallani_2007}, where two laser beams with wave lengths
$\lambda_1=830$ nm and $\lambda_2=1\,076$ nm
{are superposed.} Since these two wave lengths are highly incommensurable the disorder can be regarded as truly random, see Ref.~\onlinecite{aspect_anderson_2009}.
Alternatively,  disorder
can be  generated by superposing a laser speckle
field~\cite{clement_2006,Zhou_2010} to the regular optical lattice.
In this case, one obtains an
exponential distribution
$q(\epsilon_i) = \theta(\epsilon_i) \exp(-\epsilon_i/\bareps) / \bareps$
where $\bareps$ specifies the mean disorder strength \cite{clement_2006,Zhou_2010}.
In our calculations the distribution is additionally shifted about its median $\bareps\, \ln 2$
to avoid modifications of the chemical potential due to disorder.
The exponential distribution
is asymmetric and unbounded and thus, strictly speaking, the Mott
phase, which is controlled by the extrema of the
distribution \cite{Freericks_1996,Svistunov_1996,Pollet_2009,Gurarie_2009}
does not occur anymore.
However, the goal of the present paper is to mimic the results of the experimental measurements,  for
which the effects of the tail of the distribution are too small to be observable. Therefore
we introduce a cut off for the disorder
distribution of $4\,\bareps\,\ln 2 $. In other words, realizations far off the median are
not considered.

%sf1

\section{\label{sec:method}Variational cluster approach}
%\textit{Variational cluster approach.---}
We employ the variational
cluster approach (VCA) to investigate the phase boundary and spectral
properties of the disordered BH model. VCA has been originally
proposed for fermionic systems by M.~Potthoff \textit{et al.} in
Ref.~\cite{potthoff_variational_2003} and has been extended to bosonic
systems in
{Refs.~\cite{koller_variational_2006,aichhorn_quantum_2008,knap_spectral_2010}.}
The mathematical concept for the treatment of
disordered systems by means of VCA can be found in
Ref.~\cite{potthoff_self-energy_2007}.
In this paper we present the first application of VCA for disordered systems and
on top of that, as a new contribution {to VCA theory,} we
 extend the so-called $\mat Q$-matrix formalism
\cite{aichhorn_variational_2006,zacher_evolution_2002,knap_spectral_2010}
to a formulation suitable for disordered systems.

The basic concept of VCA is that
the exact self-energy
$\mat \Sigma_{ex}$
of a certain interacting system is given by the saddle point of a functional $\Omega[\mat \Sigma]$, which
  becomes the exact grand potential at
$\mat \Sigma=\mat
\Sigma_{ex}$~\cite{potthoff_self-energy-functional_2003-1,potthoff_self-energy-functional_2003}.
The important point is that the functional $\Omega$ can be computed exactly for a
restricted set of self-energies associated with a
so-called reference system, described by an Hamiltonian $\hat H'$,
which shares
  its interaction part with the physical system.
Typically, the reference
system $\hat H'$ is a decomposition of the physical system into
independent clusters of {a given} size $L$, so that
  its  Green's function can be determined numerically.

Disorder is treated by introducing a reference system which shares, in
  addition to the interaction part, the disorder distribution with the physical system.
The expression for the averaged grand-potential within VCA then reads
 \cite{potthoff_self-energy_2007}:
\begin{equation}
 \Omega_p = \Omega^\prime_p  +
 \Tr \ln[ - \mat{G}_p^{-1}]-
  \Tr \ln[ - \mat{G}_p'^{-1}]\;,
 \label{eq:omP}
\end{equation}
where $\Omega^\prime_p$ and $\mat{G}_p^{\prime}$ are the exact disorder
  averaged grand potential and Green's function of the reference
  system, respectively, which
can be easily evaluated numerically with high accuracy.
The averaged
Green's function of the physical
  system reads
$
\mat{G}_p =  \left(\mat{G}_p^{\prime -1} -\mat V\right)^{-1}
$, where
$
\mat V \equiv \mat{G}_0^{\prime -1}-\mat{G}_0^{-1}
$,
and
$\mat{G}_0$ and $\mat{G}_0'$
 are the noninteracting Green's functions
of the pure physical and reference systems, respectively.

{In Ref.~\cite{knap_spectral_2010} it has been discussed for a pure
  system that it is expedient for the evaluation of the traces in
  \eq{eq:omP} to use the $\mat Q$-matrix 
  formalism. This formalism
is based on the Lehmann representation of the Green's function
{$\mat{G}^\prime$}
of the reference system and that of the physical system
{$\mat{G}$}. In the present paper, we extend the $\mat Q$-matrix formalism to the case of disorder. 
For a specific disorder configuration $\eta$
}
{we have}
\begin{equation}
 \mat G^\prime_\eta = \mat{Q}_\eta\,\mat{g}^\prime_\eta\,\mat{S}\,\mat{Q}^\dagger_\eta\;\text{,}
 \label{eq:gEta}
\end{equation}
where ${\mat{g}^\prime_\eta}^{-1} = \omega - \mat
\Lambda_\eta$ and $({\mat \Lambda_\eta})_{rr^\prime} =
\lambda_r^\eta\,\delta_{rr^\prime}$ are the poles of the reference
Green's function, see Ref.~\cite{knap_spectral_2010} for details.
In praxis a specific number $M$ of disorder configurations $\eta$ (each consisting of $L$ disorder realizations $\epsilon_i$) has to be sampled to compute
the average of a quantity $X_\eta$ leading to $X_p \equiv \langle X \rangle_p \approx 1/M \sum_\alpha X_{\eta_\alpha}$.
The averaged Green's function $\mat G_p^\prime$ thus reads
\[
 \mat G_p^\prime = \frac{1}{M} \sum_\alpha \mat G^\prime_{\eta_\alpha} = \frac{1}{M} \sum_\alpha  \mat{Q}_{\eta_\alpha}\mat{g}^\prime_{\eta_\alpha}\,\mat{S}\,\mat{Q}^\dagger_{\eta_\alpha}\;\text{.}
\]
\begin{figure}
        \centering
        \includegraphics[width=0.48\textwidth]{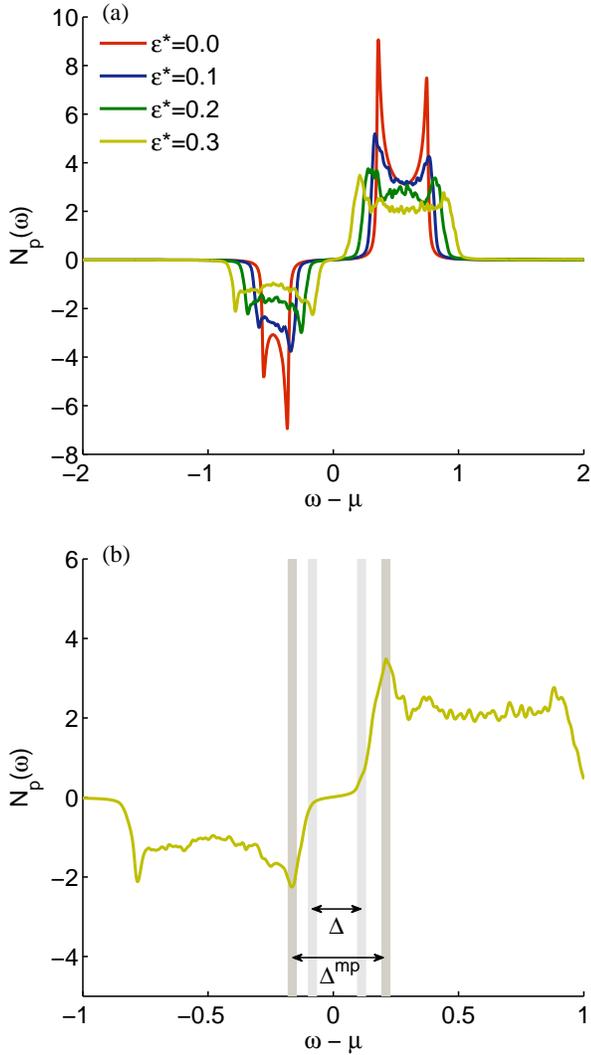}
        \caption{(Color online) \fc{a} Density of states $N_p(\omega)$ evaluated for the same parameters as
in \fig{fig:sf}. \fc{b} Illustration of the two ``gaps'' $\Delta$ and $\Delta^{mp}$
within a
blowup of the
the data for disorder strength  $\epsilon^*=0.3$.
According to our definition (see text),
$\Delta$ corresponds to the formal definition of the Mott gap based on minimal excitation energies, whereas $\Delta^{mp}$ corresponds to the minimum peak distance in the spectral weight, which is the quantity available from experiments. }%
        \label{fig:dos}
\end{figure}
To exploit the $\mat Q$-matrix formalism we write the
  averaged Green's function in a form {similar to} \eq{eq:gEta} and
  define
\[
 \mat G^\prime_p \equiv \mat{\tilde Q}\,\mat{\tilde g}^\prime\,\mat{\tilde S}\,\mat{\tilde Q}^\dagger\;\text{,}
 %\label{eq:gRefP}
\]
where $\mat{\tilde Q} \equiv ( \mat{Q_{\eta_1}}/\sqrt{M},\, \mat{Q_{\eta_2}}/\sqrt{M}\, \ldots ) $,
$\mat{\tilde{S}} \equiv \diag (\mat S,\,\mat S,\,\ldots)$ and
$\mat{\tilde g}^\prime \equiv
\diag(\mat{g}^\prime_{\eta_1},\,\mat{g}^\prime_{\eta_2},\,\ldots)$.
With that
we can proceed in the same way as for pure systems and write
the averaged Green's function of the physical system as
\[
 \mat G_p = \mat{\tilde Q}\,\frac{1}{\omega - (\tilde{\mat \Lambda} - \mat{\tilde S} \,\mat{\tilde Q}^\dagger \, \mat V \,\mat{\tilde Q} )} \,\mat{\tilde{S}} \, \mat{\tilde Q}^\dagger \;\text{,}
 %\label{eq:gP}
\]
where $\tilde{\mat \Lambda} \equiv \diag (\mat \Lambda_{\eta_1},\,\mat
\Lambda_{\eta_2} \ldots)$.
 By diagonalizing the matrix $\mat M \equiv \tilde
{\mat \Lambda} - \mat{\tilde S} \,\mat{\tilde Q}^\dagger \, \mat V
\,\mat{\tilde Q} = \mat X \, \mat D \, \mat X^{-1} $ we obtain the
poles $\mat D$ of the physical Green's function and are thus able to
evaluate the grand potential $\Omega_p(\mat x)$.

\section{\label{sec:results}Results}
%dos
%\textit{Results.---}
We first investigate spectral
properties for disorder realizations sampled from the shifted
$\beta$ distribution, which is realized experimentally by superposing
two incommensurate optical lattices, see Appendix~\ref{app:a} for details on the distribution. 
\begin{figure}
        \centering
        \includegraphics[width=0.48\textwidth]{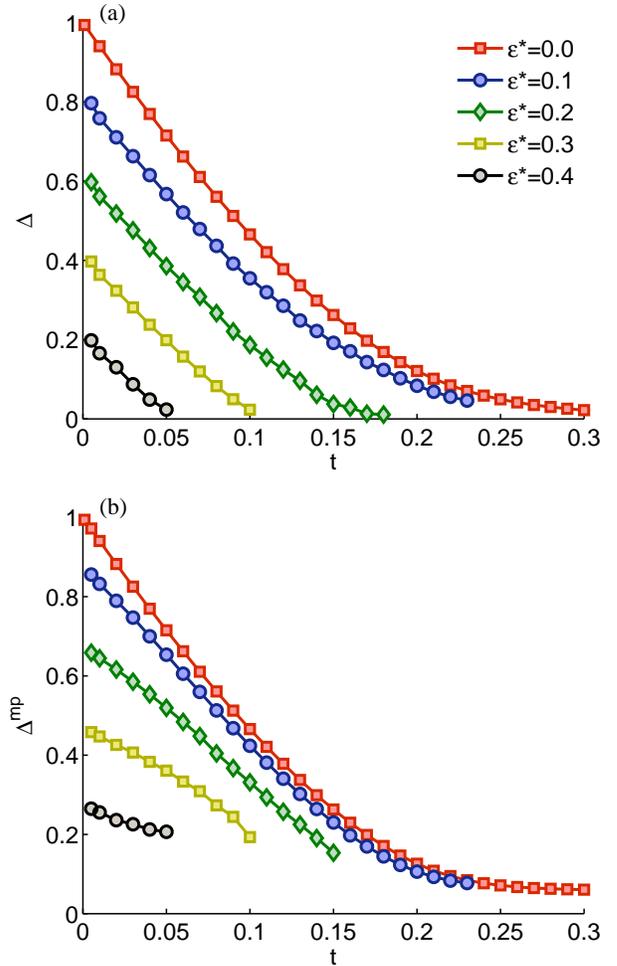}
        \caption{(Color online) Gap $\Delta$ obtained from \fc{a} the smallest excitation energies and  \fc{b} $\Delta^{mp}$ obtained from the minimum peak energy difference of the occupied and unoccupied bands.}
        \label{fig:gap}
\end{figure}
 
In particular, spectral functions
$A_p(\ve k,\,\omega)$ are evaluated for hopping strength $t=0.05$,
chemical potential $\mu=0.45$ and various strengths of disorder
$\epsilon^*=\lbrace 0.0,\,0.1,\,0.2,\,0.3 \rbrace$, see \fig{fig:sf}.
{The corresponding densities of states $N_p(\omega)$ are shown in \fig{fig:dos}.}

For the numerical evaluation by means of VCA we use
$M=256$ disorder configurations, a reference system of size $L=8$, and
the VCA parameters $\mat x = \lbrace \mu,\,t,\,\delta
\rbrace$. The parameter {$\delta$} is an additional on-site
energy located at the {boundaries}  of the cluster, {whose introduction}
drastically improves the results, as shown in Ref.~\cite{knap_benchmarkingvariational_2010}.
The artificial broadening parameter is chosen to be $0^+=0.05$ for
spectral functions and $0^+=0.01$ for densities of states. As
mentioned in the introduction the  {occupied part} of the single-particle
spectral function is experimentally accessible by high-energy Bragg
spectroscopy. For the pure system we obtain the well-known cosinelike
shaped bands reminiscent of the dispersion of free particles on a
lattice.
For increasing disorder strength $\epsilon^*$ the minimal gap between the
{occupied and the unoccupied} band shrinks.
{In addition,}
the bands become broader as
the poles of the Green's function for a specific wave vector $\ve k$
are  distributed over a large energy range.
This  behavior can also be seen
in the density of states.
{Furthermore, for large disorder the bands seem to split in two
  sections separated by a pseudogap around $k=\pi$.
This is a peculiarity of this disorder configuration which to some extend
resembles a binary distribution.
An ordered binary distribution would double the unit cell thus
producing a true gap at $k=\pi$.
}

Experimentally the gap $\Delta$ present in Mott phase can be
determined for instance by lattice modulation \cite{Stoeferele_2004}
or by Bragg spectroscopy \cite{clement_2009}. In these experiments the
amount of energy transferred to the system is related to the width of
the central peak observed in time of flight images
\cite{Stoeferele_2004}. The width of the central peak is measured for
various  energies leading to the {excitation} spectrum. Indeed for
increasing disorder strength $\epsilon^*$ a broadening of the excitation
band has been observed \cite{Fallani_2007} which is qualitatively in
agreement with our results for the spectral function. In addition, it
could be shown experimentally that the weight moves to lower
excitation energies for increasing $\epsilon^*$
\cite{Fallani_2007}. However, it is rather difficult to extract the
precise value of the gap from the measured excitation spectra.
Strictly speaking,
 the gap $\Delta$ is defined as the energy difference
of the
lowest lying poles in the occupied and unoccupied bands, respectively. Yet,
our results show, that these poles
{quite generally}
carry only very little spectral
weight. 
This means that it is virtually impossible to detect them in the
experiment. {For this reason,} it might be useful to introduce {the notion
of} a gap $\Delta^{mp}$, that corresponds to the experimental situation and  
is determined by the distance between the maxima of
the spectral weight observed in the center of the Brillouin zone. In
\fig{fig:gap} we compare $\Delta$ [\figc{fig:gap}{a}] {with} $\Delta^{mp}$ [\figc{fig:gap}{b}] which is {obviously} always larger than $\Delta$.
For increasing hopping strength $t$ the gap $\Delta^{mp}$ {decreases},
however, it still remains finite (\ie, $\Delta^{mp}\approx
0.2$) 
{at values of $t$ for which}
 the gap $\Delta$ determined from the smallest excitation
energies is already almost zero. Therefore, a peak at finite energy
will be observed in the experimental data, even when the system is
already {in the} Bose-glass phase. Additionally, it
is important to mention that the smallest excitation energy gap
$\Delta$ can be {predicted} analytically for systems with infinitely
many disorder realizations. In this case $\Delta$ is
{controlled}
by the maximum disorder strength $\epsilon^*$ {only}.
In particular, the
{phase boundary} is shifted by $\pm \epsilon^*$ as there exist
always rare regions where the chemical potential is either increased
or decreased by $\epsilon^*$
\cite{Freericks_1996,Svistunov_1996}.
Here, however, we
take into account 
a finite number of $L\,M=2\,048$ random values for the on-site energies $\epsilon_i$
leading to larger gaps, since
{it is very unlikely that all values $\epsilon_i$ of one realization $\eta$ are close to}
the extreme cases {$ \pm \epsilon^*$}.
This resembles more closely the experimental situation in
which only a limited number of disorder realizations can be detected.

\begin{figure}
        \centering
        \includegraphics[width=0.48\textwidth]{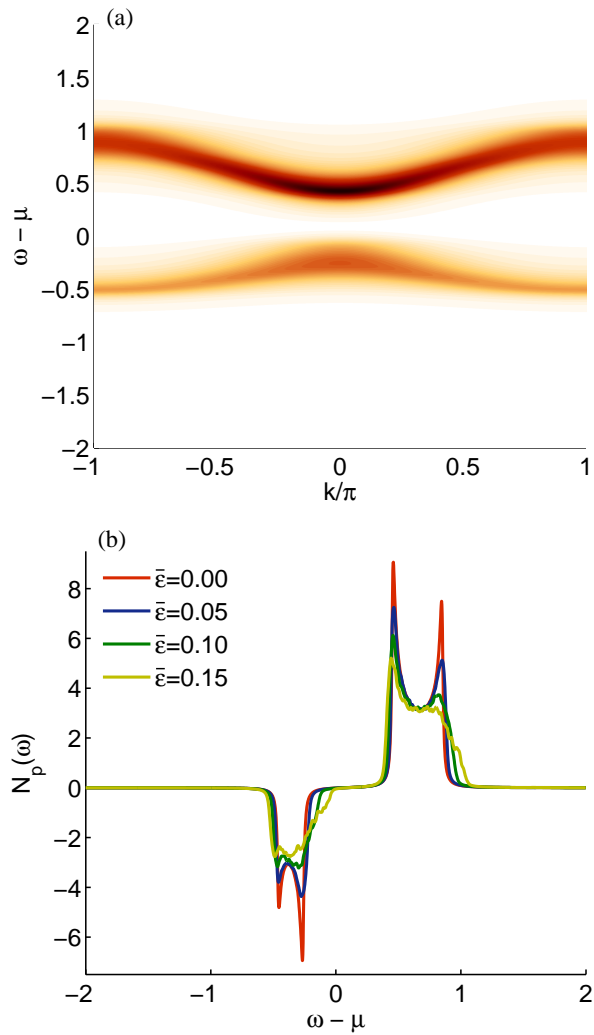}
        \caption{(Color online) Spectral properties for
a disorder distribution generated by laser speckle fields. The
parameters are $t=0.05$, and $\mu=0.35$. Panel \fc{a} shows the
spectral function $A_p(\ve k,\,\omega)$ evaluated for $\bareps=0.15$
and panel \fc{b} shows the densities of states for various disorder
strengths $\bareps=\lbrace 0.0, \,0.05, \,0.1,\,0.15 \rbrace$.}
        \label{fig:sfExp}
\end{figure}

The second kind of disorder we are addressing in this paper follows
the shifted exponential distribution, which is generated by
superposing a laser
speckle field. Spectral properties
{of the Bose-Hubbard model}
for this disorder distribution are
shown in \fig{fig:sfExp}.
%%
%sfExp
In contrast to the disorder realized by {the superposition of two} incommensurate
optical lattices, the spectral signatures evaluated for the exponential
distribution clearly {exhibits} an asymmetric shape. This becomes
particular visible when comparing the densities of states for various
disorder strengths, see \figc{fig:sfExp}{b}. In particular, the poles are smeared out toward higher excitation energies and thus reflect the tail of the exponential distribution. No pseudogap behavior around $k=\pi$ is observed here, due to the shape of the exponential distribution which is in contrast to the $\beta$ distribution not peaked at the edges.

\section{\label{sec:conclusion}Conclusions}
%\textit{Conclusions.---}
In the present work we investigated for the
first time spectral properties of the disordered Bose-Hubbard
model.
{In view of a realistic description of the experimental results,}
we focused on
disorder distributions which are relevant for
ultracold gases of atoms in optical lattices. In
particular, we
{studied the differences between}
disorder realized by {the
  superposition of }
incommensurate  {laser fields with the one obtained by
 laser speckle fields.} In both
cases we evaluated spectral functions and densities of states and
showed that the resulting spectral weight  {strongly depends} on the
underlying shape of the disorder distribution. Furthermore, we
determined the gap present in the Mott phase for disorder generated by
incommensurate optical lattices. On the one hand, we evaluated the gap
$\Delta$ from the minimal excitation energies of the system and on the
other hand we {determined} the gap $\Delta^{mp}$ from the minimum peak
distance in the spectral weight located at the center of the Brillouin
zone. Whereas $\Delta$ cannot be observed directly in the experiment since
the low-energy excitations carry very little spectral weight,
$\Delta^{mp}$ is directly measurable. Furthermore,
$\Delta^{mp}$ is always larger than $\Delta$ and thus $\Delta^{mp}$
{remains finite }
even at the Mott to Bose-glass transition. 
In our calculations, we neglected the harmonic trap potential
 present in the experiments with ultracold gases of atoms. 
In principle, this effect can be included in our formalism, however,
with a significantly major effort which goes beyond the goal of the
present work.
There are two cases in which neglecting the trap potential is justified.
As in the ordered case,
one can expect for sufficiently small disorder strength
the trap potential to lead to
multiple ringlike regions which are alternately Mott gapped and
gapless.
Quite generally, one should be able to choose the parameters so that the volume of
the gapless regions is much smaller than the one of the $\hat N_p=N$ 
Mott region  in which we are interested.
In this case, the spectrum will display a nonvanishing weight within the gap
originating from the gapless regions. However, we expect this to be
small enough,
 at least for $\omega \not=0$, so that the
peaks defining the experimental gap $\Delta^{mp}$ remain discernible.
Alternatively, spectroscopy experiments probing the system locally 
should be able to probe directly the Mott insulating region with no
contributions from the gapless areas.

\begin{acknowledgments}
M.K. thanks P. Pippan for many useful discussions. We made use of
parts of the ALPS library \cite{albuquerque_alps_2007}.
We acknowledge financial support from the Austrian Science Fund (FWF)
under the doctoral program ``Numerical Simulations in Technical
Sciences'' Grant No. W1208-N18 (M.K.) and under Project No. P18551-N16
(E.A.).
\end{acknowledgments}

\appendix
\section{\label{app:a} Disorder distribution generated by incommensurate optical lattices}
In this Appendix we show that the potential distribution, which results from the superposition of two optical lattices, follows a shifted $\beta$ distribution. In particular, we focus on the experimental setup of Fallani \textit{et al.} \cite{Fallani_2007}, who used for the main optical lattice a laser at wavelength $\lambda_1 = 830$ nm. Disorder is generated by superposing an additional lattice created from a weak laser beam at $\lambda_2=1\,076$ nm. The resulting potential is given by
\[
 V(x) = s_1\,E_{R1} \sin^2 2\pi x/\lambda_1 + s_2\,E_{R2} \sin^2 2\pi x/\lambda_2 \;\mbox{,}
\]
where $x$ is the spatial position, $s_1$ and $s_2$ are related to the depth of the potential generated from the laser beams at $\lambda_1$ and $\lambda_2$, respectively. The constants $E_{R1}$ and $E_{R2}$ are the corresponding recoil energies. In this Appendix, the lattice depths $s_1$ and $s_2$ will always be stated in units of their recoil energies. The depth $s_2$ of the disorder-inducing wave is related to the maximal disorder strength $\epsilon^*$ by $\epsilon^*=s_2/2$. Since the wave lengths $\lambda_1$ and $\lambda_2$ are incommensurable the disorder imitates a true random behavior \cite{aspect_anderson_2009}. 
Here we reconstruct the histogram of the disorder distribution and find a mathematical expression, which reproduces the behavior of these physical systems.
\begin{figure}
        \centering
        \includegraphics[width=0.4\textwidth]{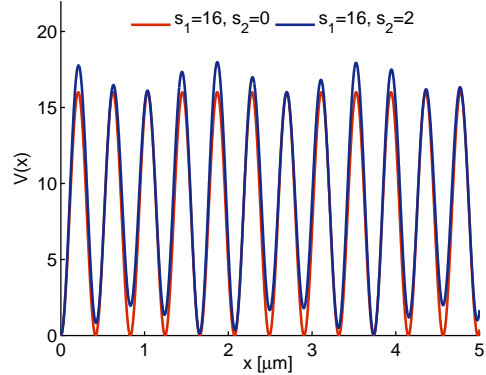}
        \caption{(Color online) Total lattice potential created by the superposition of the main optical lattice with depth $s_1$ and the disorder lattice with depth $s_2$. }
        \label{fig:potSupp}
\end{figure}

\begin{figure}
        \centering
        \includegraphics[width=0.4\textwidth]{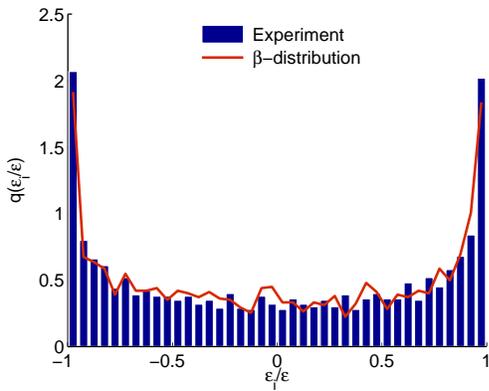}
        \caption{(Color online) Distribution of the on-site energy $\epsilon_i$ observed in the experiment (histogram) compared with random samples drawn from the shifted $\beta$ distribution (solid line). }
        \label{fig:betaDist}
\end{figure}

In the experiments typical values for the lattice depths are $s_1=16$ and $s_2=2$, see Ref.~\cite{Fallani_2007}. \Fig{fig:potSupp} compares the potential $V(x)$ for the previously mentioned values of $s_1=16$ and $s_2=2$ with the pure case, where the second laser beam at $\lambda_2$ is switched off, \ie, $s_1=16$ and $s_2=0$.
%potSupp
It can be seen that the on-site energy varies for distinct lattice sites mimicking a random potential. Actually, the parameter $s_2$ just scales the disorder and thus our considerations are valid for arbitrary disorder strength $\epsilon^*$. To evaluate a histogram of the energy distribution we subtract the disordered potential at the lattice sites from the pure potential and shift the difference by its mean $\epsilon^*$. This yields a distribution which is centered around zero. The shift could have been absorbed as well in the definition of the chemical potential $\mu$. The histogram for $2\,048$ lattice sites is shown in \fig{fig:betaDist}. 
%betaDist
In the center the distribution is rather flat, yet, there are important features at the boundary. Such a distribution can be well described by a shifted $\beta$ distribution. 

The $\beta$ distribution $q_{\beta}(u|a,\,b)$, which is defined on the interval $u \in [0,\,1]$, is given by
\[
q_{\beta}(u|a,\,b) \equiv \frac{1}{B(a,\,b)} \, u^{a-1} \, (1-u)^{b-1}\,\mbox{,} 
\]
where 
\[
B(a,\,b) = \int_0^1 \,dp \, p^{a-1} (1-p)^{b-1} = \frac{\Gamma{(a)}\Gamma{(b)}}{\Gamma{(a+b)}}\,\mbox{.}
\]
The expectation value of $q_{\beta}(u|a,\,b)$ is 
\begin{equation}
 \langle u \rangle  = \frac{a}{a+b}
 \label{eq:mean}
\end{equation}
and its variance is 
\begin{equation}
 \text{var}(u) = \frac{\langle u \rangle (1-\langle u \rangle)}{a+b+1}\,\mbox{.}
 \label{eq:var}
\end{equation}
To obtain a distribution which is symmetric around zero and bounded by $[-1,\,1]$ we set $\langle u \rangle = 0.5$, shift the whole $\beta$ distribution by this value and scale it by a factor of 2. Resolving \eqq{eq:mean}{eq:var} under the condition that $\langle u \rangle = 0.5$ leads to $a = b = [\text{var}(u)/4 -1]/2$. For $\text{var}(u)>1/12$ (where $1/12$ is the variance of the uniform distribution), the probability density is shifted toward the boundaries of the distribution. In particular we set $\text{var}(u) = 0.12$. Finally, we draw $2\,048$ samples from this specific shifted $\beta$ distribution, which we denote as $q(\epsilon_i/\epsilon^*)$. The resulting distribution of the $2\,048$ samples is indicated by the solid line in \fig{fig:betaDist}, which reproduces well the distribution obtained from the experiment with two incommensurate optical lattices.

\end{document}